# The human genome and drug discovery after a decade. Roads (still) not taken

Ruth Isserlin[1] , Gary D. Bader[1], Aled Edwards[2] *, Stephen Frye[3], Timothy Willson[4], Frank H. Yu[5]

[1] The Donnelly Centre, Faculty of Medicine, University of Toronto, 160 College Street, Toronto, Ontario, Canada M5S 3E1
[2]University of Toronto, Toronto, Ontario Canada, M5G 1L7
[3]Center for Integrative Chemical Biology and Drug Discovery, Eshelman School of Pharmacy, University of North Carolina at Chapel Hill, Chapel Hill, North Carolina, USA 27599
[4]GlaxoSmithKline, Research Triangle Park, North Caroline, USA 27709
[5]Neurobiology and Dental Research Institute, School of Dentistry, Seoul National University, Seoul, South Korea, 110-749

* Corresponding authors: Aled Edwards – E-mail: aled.edwards@utoronto.ca

## Abstract

The draft sequence of the human genome became available almost a decade ago but the encoded proteome is not being explored to its fullest. Our bibliometric analysis of several large protein families, including those known to be “druggable”, reveals that, even today, most papers focus on proteins that were known prior to 2000. It is evident that one or more aspects of the biomedical research system severely limits the exploration of the proteins in the 'dark matter' of the proteome, despite unbiased genetic approaches that have pointed to their functional relevance. It is perhaps not surprising that relatively few genome-derived targets have led to approved drugs.

## The Harlow-Knapp Effect

The first significant analyses of the citation activity for human proteins were done for the human protein kinome by Harlow (Grueneberg et al. Proc Natl Acad Sci U S A 2008), and then by Knapp (Fedorov et al. Nat Chem Biol 2010). The human protein kinome comprises 518 members and has been the subject of intense research activity in the public and private sectors (Grant Cell Mol Life Sci 2009). Harlow and colleagues highlighted that a relatively small set of kinases garnered the majority of citations, and that most essential protein kinases, as revealed by RNAi studies, remain relatively unexplored in the scientific literature (Grueneberg et al. Proc Natl Acad Sci U S A 2008). Knapp and colleagues extended this observation to drug discovery by showing that the same small fraction of the human protein kinases that garnered the most citations were also the focus of most patents (Fedorov et al. Nat Chem Biol 2010). We thus define the Harlow-Knapp (H-K) effect as the propensity of the biomedical and pharmaceutical research communities to focus their activities, as quantified by the number of publications and patents, on a small fraction of the proteome.

The existence of the H-K effect within the human kinome may derive from a number of causes, including explicit or subliminal pressure within the biomedical enterprise to focus in areas of communal interest, the order of discovery of a protein/gene, and/or the availability of tools to study the proteins. If one of these reason(s) was the underlying cause, the phenomenon would hold for more than the human kinome and it would be resistant to change over time. In this article, we (1) examine the prevalence of the H-K effect; (2) examine publication patterns to explore the H-K effect as a function of time; (3) explore the hypothesis that the availability of reagents is one of the mediators of the H-K effect; and (4) make recommendations as to how researchers and funders might respond in lieu of these findings, particularly given the expectations of the public that the human genome project will facilitate the development of new medicines.

<u>Publication patterns pre- and post- the availability of the human genome</u>

The H-K effect was discovered by analyzing the human protein kinome. To examine whether the H-K effect is more broadly observable, we analyzed the citation patterns for the nuclear hormone receptors (NR), which are the targets for many marketed drugs, and found that most of the research activity in this family is also focused on 10% of the family members (Figure 1). Similar patterns are observed for other “druggable” human protein families, such as the human methyltransferases and bromodomain-containing proteins (Supplemental material). We conclude that the H-K effect may be a fundamental feature of research activity and not a protein family-specific phenomenon.

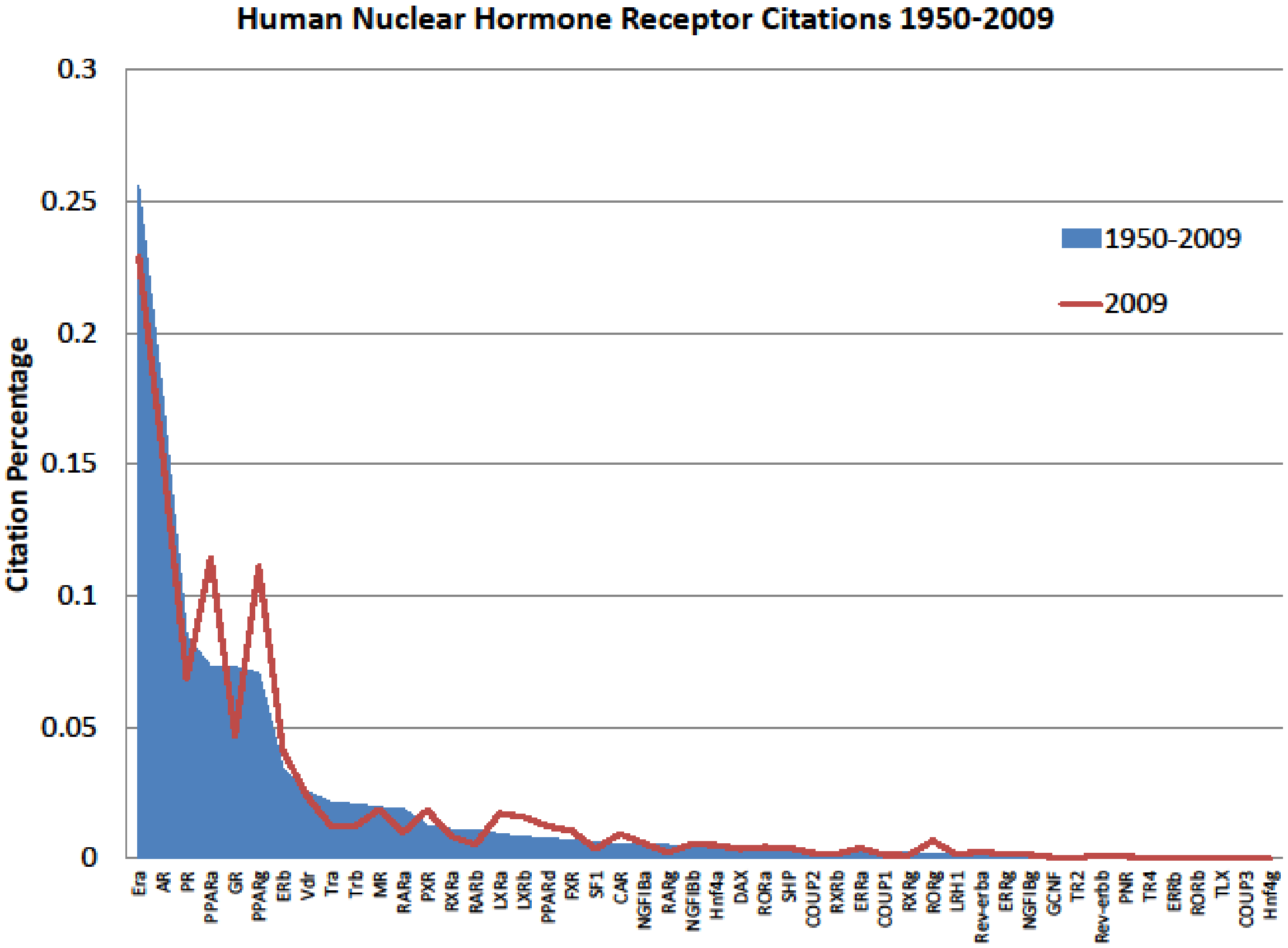

**Figure 1: Human NR Citations (1950-2009)**
Percentage of citations for 48 human NRs from 1950-2009 (116,552 total citations), ordered by total citation count over that time frame (blue). Percentage of citations for the year 2009 (11,140 total citations) is shown as a red line highlighting that the bulk of citations from 2009 is still dominated by historically popular receptors.

<u>How quickly are publication patterns changing?</u>

The H-K effect was initially seen by examining the cumulative number of citations over time (Grueneberg et al.; Federov et al.) for the set of protein kinases. It is possible that the observed H-K effect simply reflects a “carry-over” from pre-genome citations, and that present-day citation patterns are actually changing substantially to focus on the newly discovered family members. To test this idea, we performed a similar citation analysis of the protein kinase family but expanded it to include the distribution of citations overtime time. For each of the protein kinases, we quantified their citations before and after 2002, when the seminal paper charting the human kinome was published (Manning et al. Science 2002). There were a total of ~80,000 citations prior to 2002 and ~120,000 citations after. The normalized distribution of these citations is shown in Figure 2. This analysis revealed that 84% of the citations to protein kinases in 1950-2002 were focused on only 10% of the kinome (50 kinases). Interestingly, the very same kinases continued to garner most of the citations even long after the genome information became widely available (77% of citations between 2003 and 2008, and 74% of the citations in 2009). At the other end of the spectrum, the set of 300 kinases (60% of the kinome) that were at the bottom of the citation list in 2002 remain poorly studied today; in aggregate, they accrued only 5% of the kinase citations in 2009. These data suggest that the availability of the genome sequence has not substantially influenced biomedical research priorities. There are, however, indications that unbiased genome-wide studies are slowly influencing publication trends. The three largest spikes in the right hand side of the 2009 citation graph are BRAF (Entrez gene id: 673) (Davies et al. Nature 2002), PTEN Induced Putative Kinase 1 (Entrez gene id : 65018 PINK1) (Valente et al. Science 2004) and Leucine-Rich Repeat Kinase 2 (Entrez gene id : 120892 LRRK2) (Zimprich et al. Neuron 2004) – kinases that have been genetically linked to human disease in the past decade.

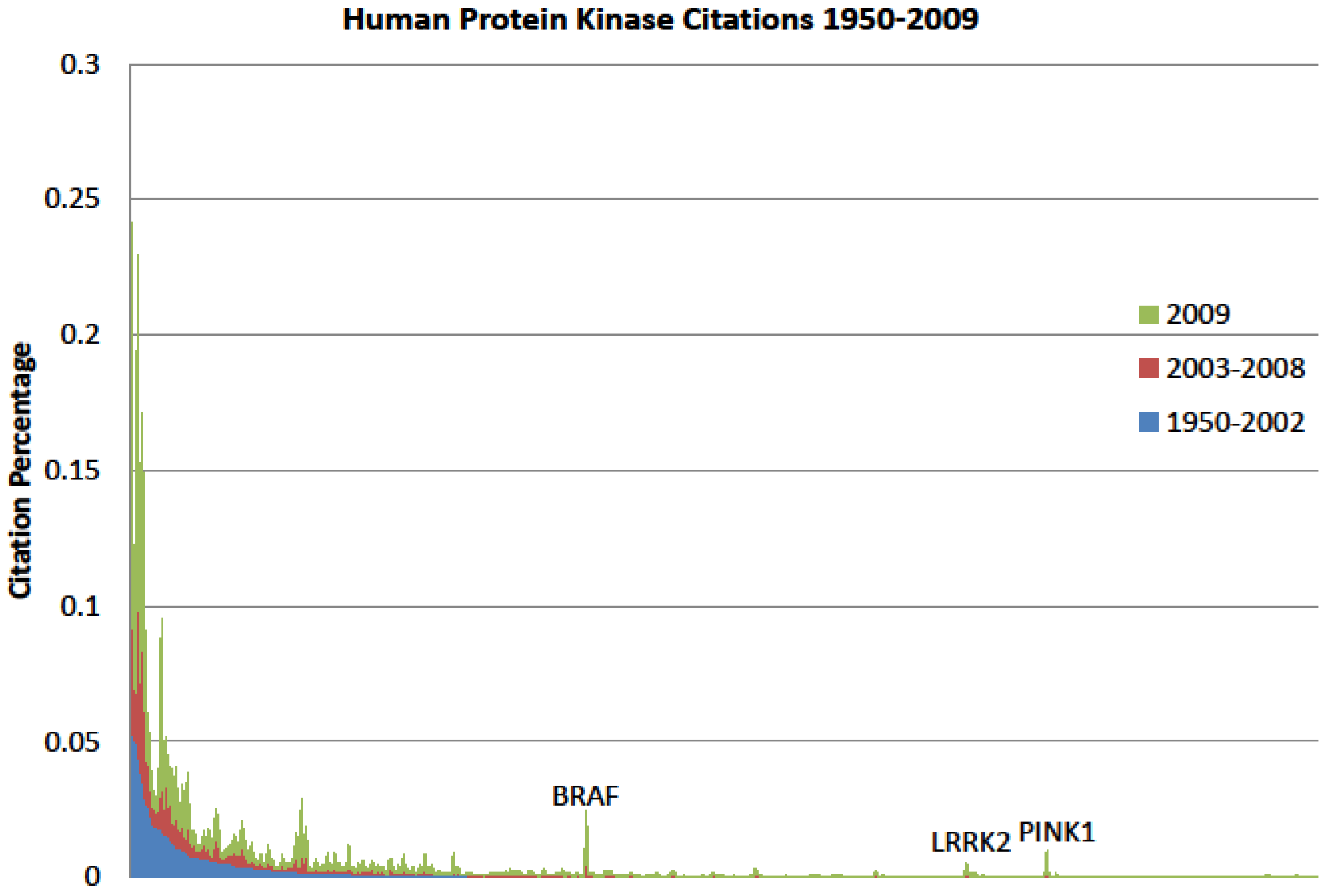

**Figure 2: Human Protein Kinase Publications 1950-2009**
Percentage of citations in the years 1950-2002 (kinome publication date), 2003-2008 and 2009 for all human protein kinases. The majority of the publications in all three year ranges, 84%, 77% and 74%, respectively, are associated with the top 50 kinases as ranked by their popularity in 2002. Exceptions to the trend include BRAF, LRRK2 and PINK1, three kinases with few publications in 2002 but substantial growth in 2009.

<u>What underlies the H-K effect?</u>

If we assume that the latest publication dealing with p38 kinase, currently one of the top ten popular kinases from our bibliometric analysis, is less impactful on our understanding of human health than the first studies of a novel kinase, then the H-K effect may reflect an underlying condition that needs intervention. Therefore, it is important to understand the basis for the H-K effect if society is to capture the most value from the human genome sequence, as well as to advance more innovative targets into drug discovery pipelines. We posit that the H-K effect likely derives from a combination of human nature, the structure of the funding and peer-review systems and the availability of research infrastructure and methods.

*Human nature.* Protein scientists often develop unique expertise within a system and then strive to attain a deeper understanding of the underlying mechanisms. Indeed, the complexity of biology is so rich that it is always possible to ask important questions about even well-studied systems, and in those efforts uncover fascinating and sometimes revolutionary biology (Guerrier-Takada et al. Cell 1983). Thus it is now well-accepted to focus on systems that are well-described, in which questions can be well-formulated and whose analysis will reveal fundamental principles that can hopefully be applied more broadly. It is also true that such discoveries can often only be made by scientists with a deep technical understanding of an experimental system. As Kornberg stated “scientists love to fondle their problems” (Judson, “The Eighth Day of Creation”, Simon and Schuster (1979)). Taken together, these considerations provide sufficient justification for grant panels to support focused science; unfortunately the corollary is that proposing to stray beyond one’s area of expertise is often criticized. In conclusion, the H-K effect derives in part from the norms of curiosity-driven science and the accepted advantages of gaining competence in a given field of research.

*Organization of funding and review.* The format of grant proposals alone encourages research within well-described areas. Grant proposals require that the applicant explain the rationale and significance of the work within the context of existing knowledge, and also to present sufficient preliminary data to convince the reviewer that the proposed research has a reasonable probability of success. These expectations and parameters strongly bias researchers to work on well-studied systems, in which peer endorsement is more likely and the path to results clearer.

Scientific success also depends on high quality publications – often assessed by the number of citations or the perceived impact of the journal. Here the system also biases toward well-studied systems because peer-reviewers of manuscripts are more comfortable if work can easily be placed into the context of existing knowledge. Compared with the analysis of a well-known protein, elucidating the function of a novel protein to a level sufficient to be accepted into a “top” journal requires the investment of significantly more effort and resources, and this is much riskier for scientists concerned with timelines of grant renewals, student theses and tenure clocks.

The organization of grant applications also conspires against research of the proteome’s dark matter. Given that the most important metrics used to score grants are: context within existing knowledge; innovation; and probability of success/impact, proposals on unstudied proteins are less competitive compared with proposals that are more well-contextualized. Indeed, proposals on unstudied proteins are often damned with the pejorative “fishing expedition” label. It is therefore quite understandable that many in the scientific community elect to focus on scientific areas endorsed by their peers.

*Lack of research tools and methods.* The availability of research tools also has a significant impact on the analysis of the human proteome. The field of genetics and genomics show that scientists by nature are not averse to unbiased genome-wide studies. Many current studies of biology exploit genome-wide tools as a matter of routine, e.g. SNP arrays or whole genome RNAi sets. However, the in-depth analysis of any protein that emerges from these screens depends on the availability of high quality, protein-specific research tools, such as the purified protein, high quality antibodies and/or high quality chemical probes. Such protein-specific reagents are not as readily available, and thus it often falls to the investigator to develop the enabling tools. For any laboratory or field, the generation of these tools presents a major and risky commitment of time and funds (http://www.biotechniques.com/news/Turning-the-tables-on-antibody-validation/biotechniques-300587.html). When the protein is directly linked to a large disease, such as cancer or neurological disease, has a predicted biochemical function, and has "druggable" properties, the risk appears worth the reward, as evidenced by the rapid growth in publications for the disease-linked BRAF, PINK1 and LRRK2 kinases. However, if a protein does not have these favourable features, it is more difficult to make the case to take the time to generate the necessary research tools. As a result, in most instances when presented with a range of "hits" from genome-wide genetic screens, scientists are compelled to pursue the proteins for which there already exist useful reagents - which *de facto* are already among those most well-studied - and set aside the proteins of unknown function for which there are no research tools.

## How to change the system?

Granting agencies are certainly aware of the difficulty in supporting innovative research, and many have explored and are continuing to explore innovative models to fund riskier science. The struggle in this effort stems from the peer-review system, which endorses science within accepted bounds and is inherently risk-averse, and from scientists themselves whose careers depend on performing within the system and within strict time frames.

In this paper, we will explore the role that the availability of research tools plays in the biomedical research endeavour and will conclude that one way to encourage a broader range of studies down "roads not taken" would be to generate sets of high-quality reagents for all human proteins.

<u>The availability of research tools and publication patterns</u>

Genetic studies have linked thousands of human genes to interesting phenotypes, yet the publication records show that only some of the interesting proteins are pursued vigorously by the community. It is likely that the choices about which proteins receive attention are in part influenced by the availability of reagents for each protein.

We set out to test this idea by examining the retrospective publication trends in the 48-member human NR family. NRs are transcription factors that directly bind small molecule signaling ligands, such as steroids, hormones and cholesterol derivatives. This family presents an ideal test case because: (1) many of the family members were discovered at about the same time in the 1990's, allowing us to compare publication trends for a large number of related proteins; (2) when discovered, all members of the family were considered "druggable", of potential therapeutic relevance and thus "exciting", yet their biological roles and cognate ligands were unknown; (3) the research community launched "ligand discovery" efforts contemporaneously and the dates at which the resulting research tools, specifically chemical inhibitors and activators, became available are known; (5) there is now an equal number of family members for which tools are and are not available.

Industry became very interested in determining the functions of the orphan receptors because it was thought that many would be drug targets, based on the druggability of the set of receptors already known at the time. Among the strategies employed by the companies to determine their function was one termed “reverse endocrinology”, in which they attempted to develop potent and selective synthetic agonists and/or antagonists to all the receptors so as to apply the “chemical probes” to cells and animals to perturb their normal functions in cell physiology. These approaches were applied without bias, and probes were discovered for some but not all NRs. This protein family therefore provides the ideal set to analyze the effects of chemical probes on publication trends.

In analyzing the NR citation patterns over time, we found that the six NR’s (12.5% of the family) that were the top cited in 1999 (Estrogen Receptor alpha (Entrez gene id:2099, Era), Progesterone Receptor (Entrez gene id:5241, PR), Androgen Receptor (Entrez gene id:367, AR), Glucocorticoid Receptor (Entrez gene id:2908, GR), Peroxisome Proliferator-Activated Receptor alpha, and gamma (Entrez gene id:5465, PPARa, Entrez gene id:5468, PPARg) remained the top cited in 2009. These six NR’s garnered 71% of the NR citations in 1999, and a slightly greater fraction in 2009 (72%) – perhaps because there are marketed drugs to each. Interestingly, in contrast to the protein kinome, where over the past decade we observed a drifting away from the kinases of historical interest, in the NR field, this is not occurring.

To examine the impact on the availability of chemical tools on publication trends, we examined the citation rates of the orphan receptors since their discovery, and the changes in the citation patterns once tools were made available. Prior to the availability of chemical tools, the orphan receptors that most rapidly developed interest in the literature were those that had genetic links to disease, such as HNF4a, SF1 and DAX1 (Muscatelli et al. Nature 1994; Yamagata et al. Nature 1996; Achermann et al. Nat Genet 1999), or that had interesting knock-out phenotypes, such as NGF1-B (Lee et al. Science 1995) (Supplementary Figure 1). This pattern appears common to protein kinases (e.g. BRAF, PINK1 and LRRK2), NRs (e.g. HNF4) and ion channels (see below); in each family, we can see that the community’s initial interest in a protein is often spurred by a genetic link to a human disease. This observation highlights the seminal role of human genetics and genomics in capturing the interests of the scientific community. However, our analysis of the NR field suggests that, in the longer term, ongoing research activity is influenced more by the availability of research tools than by disease link.

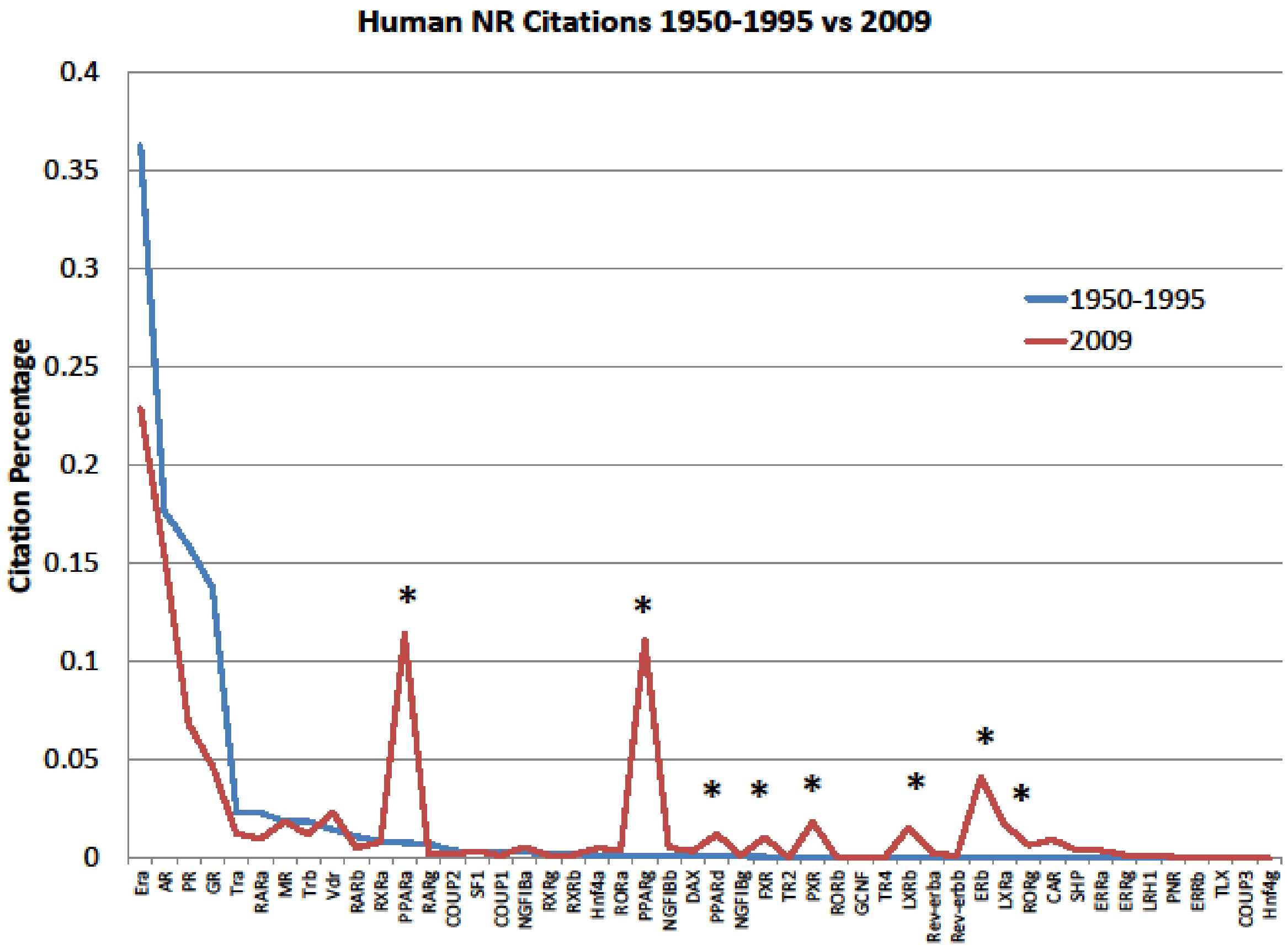

**Figure 3: Human NR Citations 1950-1995 vs. 2009**
NR citation rates for 1950-1995 (when most NRs were discovered) and 2009 order by total in 1950-1995. Starred receptors include PPARa, PPARg, PPARd, FXR, PXR, LXRb, ERb, LXRa show drastic increase in citations comparing 1995 and 2009.

Prior to 1995, the NR's that were most highly cited were those for which the natural ligands were known, such as the steroid receptors including ERa, AR, PR, and GR. These were followed by those orphan receptors with interesting disease links. However, by 2009, there was a significant change in the focus of NR publications (Figure 3); eight orphan receptors become major topics of research including PPARa, PPARg, PPARd, Farnesoid X Receptor (Entrez gene id:9971, FXR), Pregnane X Receptor (Entrez gene id:8856, PXR), Liver X Receptor alpha and beta (Entrez gene id: 10062, LXRa), (Entrez gene id:7376, LXRb), and Estrogen Receptor beta (Entrez gene id: 2100, ERb). To our knowledge, the only connection between each of these receptors is that for each there is a commercially available, high quality chemical probe. Indeed, if one analyzes the correlation between the research activity in 2009 and the availability of chemical tools, one observes that of the top 19 cited receptors each of them had chemical tools available, whereas only 2 of the bottom 29 had an available high-quality chemical tool (Figure 4). Interestingly, it is not the existence of a chemical tool that is the most important, but instead the fact that they were of high quality and commercially available. Although there are chemical tools that have been described for many of the less well-cited NRs, these molecules have either poor cellular activity, lack bioavailability or were not commercially available. (Charpentier et al. J Med Chem 1995; Buijsman et al. Med Chem Res 2004; Busch et al. J Med Chem 2004; Zuercher et al. J Med Chem 2005;

Whitby et al. J Med Chem 2006; Grant et al. ACS Chem Biol 2010; Wang et al. ACS Chem Biol 2010).

In summary, ~30 orphan receptors were co-discovered in the mid-1990's. A priori, there was no reason to rank them by popularity in the order that developed over the past 15 years. Although many are linked with interesting genetic phenotypes, the ranking that developed appears to correlate only with the availability of high quality reagents, suggesting that post-genome academic research directions are heavily influenced by the availability of research tools.

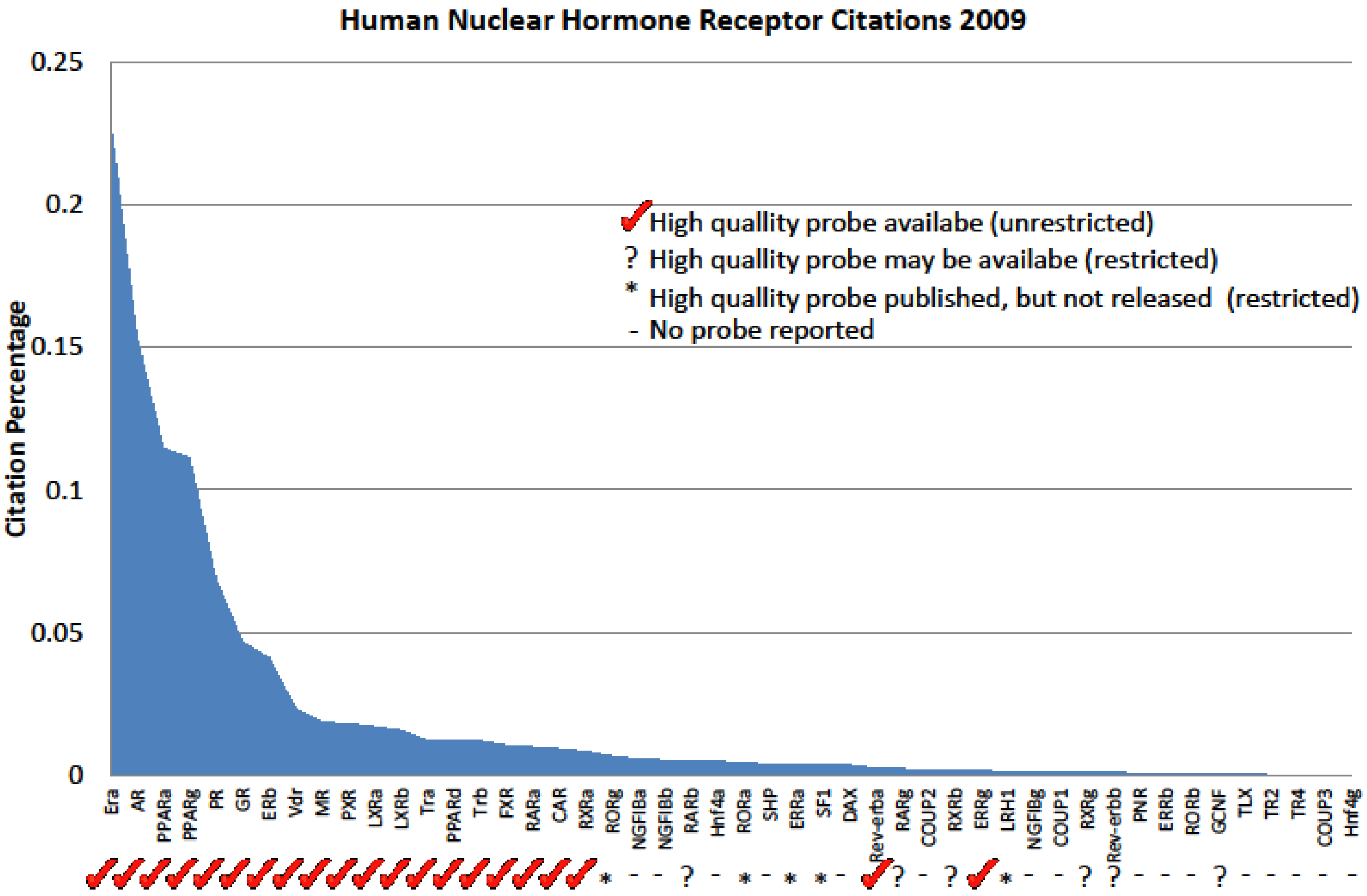

**Figure 4: Chemical Tools associated with Human Nuclear Hormone Receptors**
Percentage of citations for the 48 NRs for 2009. Receptors are annotated according to chemical tools associated with them, (red check mark) high quality probes with unrestricted availability, (?) with high quality probes might be available but are restricted, (*) with high quality probes published but not released, and (-) with no probes reported. The top cited 19 receptors have high quality probes with unrestricted availability, two exceptions are Rev-erba and ERRg that have relatively few publications in 2009 in comparison to the rest of the receptors with published chemical tools.

## Causation or correlation?

Our hypothesis is that the availability of research tools in part drives the direction of biomedical research. Although it is impossible to disentangle causation from correlation, three observations about the NR field are consistent with this hypothesis. First, if one examines the citations of each of the 48 NR's over time, one finds a remarkable temporal correlation between the release of a chemical probe and the increase in publication intensity with respect to the other NR's (Supplementary Figure 1 and 2 ). Second, the papers that describe the first chemical probe for each NR are among the most highly cited for that receptor; for example, for the PPARs, PXR, LXRs, FXR and CAR, the initial probe paper(s) remains among the top 3 in total citations for that receptor (Supplementary Table 1). In the case of PXR, now the tenth most-cited NR, the top 3 cited papers each report a chemical probe (Supplementary Table 1). Third, of the 12 most highly cited papers for NR's, two report the development of a probe and the other 10 make use of the probe in the study (Supplementary Table 2).

*Publications reporting chemical tools for protein kinases also have significant citation impact*

For NR's, we observed a strong correlation between the availability and timing of the publication of a chemical probe and the relative publication activity, and also showed that the 12 most highly-cited papers in the field either describe or exploit chemical probes. To explore whether this was a more general phenomenon, we studied the protein kinase family. First, we identified the most highly cited papers that contain the terms "protein kinase" or "tyrosine kinase" in their titles, both over all time and over the past 5 years, then examined them to find those that describe or use chemical tools and finally assessed their impact using citation analysis.

Of the most-highly cited 20 papers that contain the term "protein kinase" or "tyrosine kinase" in their titles, 7 describe the first use of a chemical inhibitor and 8 more make use of an inhibitor in the study (Supplementary Table 3). The 15 papers that exploit chemical tools garnered 73% of all the citations for the top 20; the seven probe-report papers alone captured 36% of the top 20 citations.

Chemical probes for protein kinases continue to have a major impact on the literature. We identified the Top 20 papers published from 2005 to 2010 with "protein kinase" or "tyrosine kinase" in their title and, of these, 16 make use of chemical tools – garnering 66% of the "Top 20" citations. Six of the most highly cited 20 papers over the past 5 years described the clinical use of protein kinase inhibitors. Interestingly, 3 of the 4 "Top 20" papers that did not use chemical tools were DNA sequencing papers that linked specific mutations to human cancers – demonstrating the impact of the human genome sequence.

*Ion channels. In the absence of tools, disease associations dominate.*

Ion channels represent another protein family that is the subject of intense research activity. And in this family, the H-K effect is evident; 10% of the most highly cited channels in 2009 garnered 50% of the research activity (Supplementary Table 4). One channel in particular, TRPV1, alone attracted almost 10% of the research activity in 2009. Like the protein kinases, the under-studied ion channels receive little attention. The lowest 60% of the ion channels garnered only 12% of the citations, though there are many examples of under-studied channels of great importance (Lafreniere et al. Nat Med 2010).

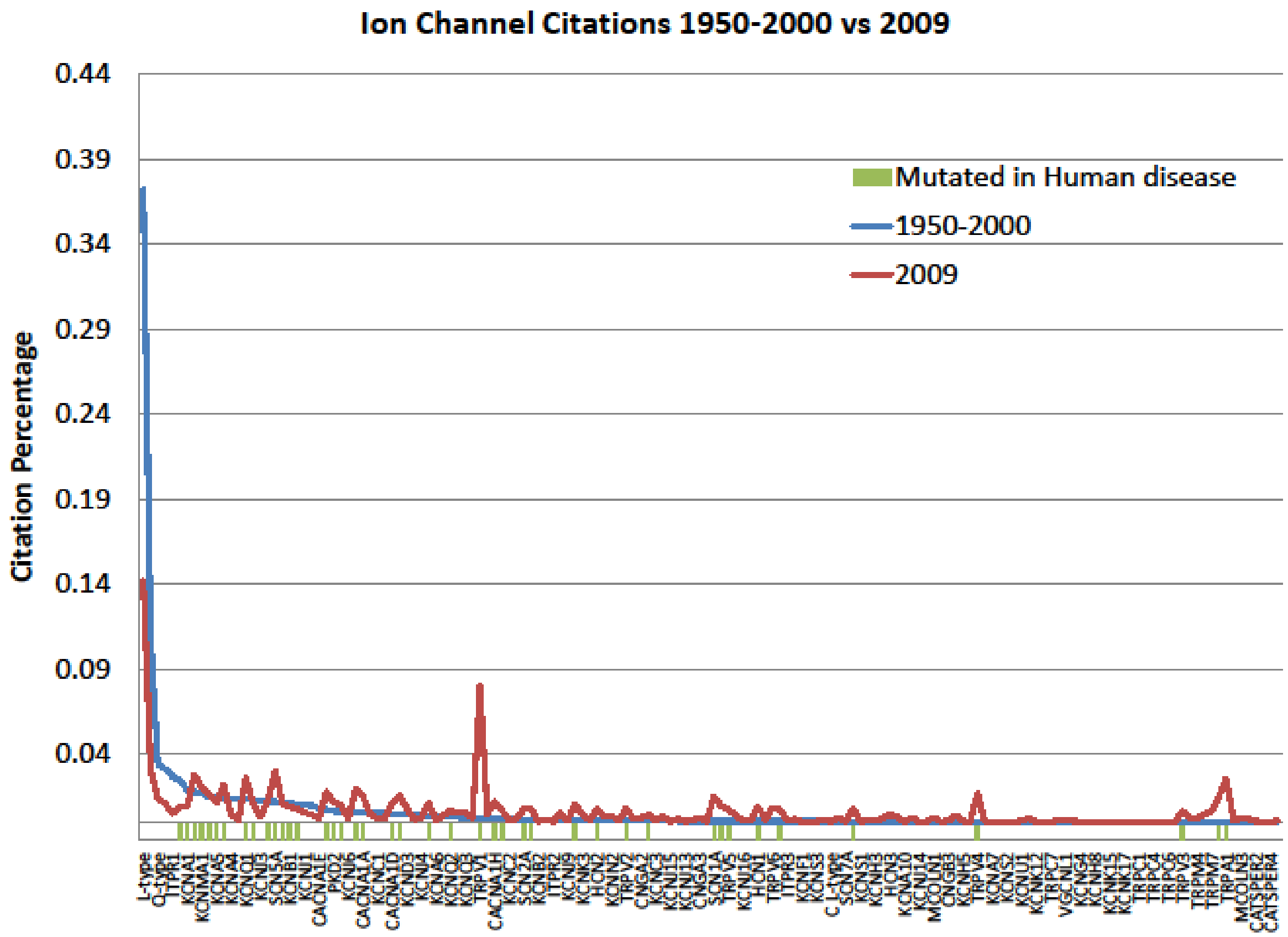

**Figure 5: Ion Channel Citations 1950-2000 vs. 2009**
Percentage of Citations for ion Channels for 1950-2000 compared to 2009. Channels are annotated if they have been shown to be mutated in human disease. Receptors associated with disease often show increase citation rate if they are associated with disease.

Within the ion channel family, there are many members that are mutated in human disease, and many others that are linked to disease as a result of their knock-out phenotype. For the ion channels, the publication trends seem to correlate well with genetic evidence for disease relevance, somewhat in contrast to the NR and protein kinase field, in which the availability of protein-selective chemical tools also had great influence. For the ion channels, this may be explicable because there are still few channel isoform-specific antagonists or agonists, either chemical or protein based. As a result, research on specific channels is carried out using molecular biology approaches, which are dependent only on the availability of the cDNA. The paucity of channel-specific probes is interesting given that research in ion channels grew rapidly after it was learned that toxins and chemicals could functionally differentiate various ion channel classes (e.g. tetrodotoxin for voltage-gated sodium channels, tetraethylammonium for voltage-gated potassium channels, $Ba^{2+}$ for voltage-gated calcium channels).

## Summary and future directions

Our bibliometric analysis of human protein families revealed that the Harlow-Knapp effect is not restricted to protein kinases and that citation trends are slow to change, likely due to inherent features of scientists and of peer-review systems. We also found that genetic linkage to disease is a major driver of research trends, but that the availability of research tools has a dominating influence. Linking a gene/protein to a disease is not always enough to change publication patterns.

On the one hand, our analysis supports what many scientists have known intuitively; that the peer-review system fosters conservative thinking. However, the analysis reveals a solution – the reticence to explore the wider aspects of the genome diminishes greatly as research tools become available. We conclude that the generation of research tools in an unbiased way should be a major objective of the research enterprise, and there should be a reward system in place for those scientists who contribute to this effort. We also conclude that to have maximal impact, the tools must be made available from commercial sources and with no restriction on use.

In summary, while funding researchers to generate deeper insights into the biology of proteins highly precedented to be important in human biology and disease is important, the current distribution of biomedical efforts, as observable in the H-K effect, is probably out of balance relative to the opportunities hidden in the 'dark matter' of the genome. Based on the trends revealed here, increased support for reagent generation in this domain may have a broad and rapid impact on translational science as well as provide the first foothold for more fundamental biological exploration.

## Methods

For each protein family, including kinases, NRs and ion channels, a list of family members consisting of either gene symbol or name was collated by experts. Using NCBI Entrez utilities each symbol was used to query the Entrez gene database to get additional symbols, aliases, and alternate names. Due to the query expansion feature in PubMed, whereby quoted terms that are not found within the database are translated to all combinations of the words in the quoted phrase, each additional name was searched separately in PubMed to ascertain whether it exists in the database as is. If the name is present in PubMed, as is, it was added to the list of query terms. The number of records returned for each of these individual searches was stored for subsequent filtering, discussed next.

In order to help reduce false positive PubMed results, additional restrictive terms were added to each query. For kinase queries 'and "kinase"' was added, for NR queries 'and "receptor"' was added, and for ion channels 'and ("channel" OR "channels" OR "currents" OR "conductance" OR "conductances")' was added.

The constructed query was subsequently restricted to each of the years of interest. The number of records returned by PubMed was collected for each member of the family, summarized and further normalized based on the total number of records for the entire family for the given time period. In order to further refine the results, initial rankings of the genes were examined by experts to gauge the validity of the results. It was found that many genes had names associated with them, such as "APE" or "MR", that caused spurious results. The number of results returned from the separate queries of each individual name in PubMed were used to highlight names that were causing these false positives. This resulted in an exclusion list of names for each of the families designed to reduce false positive results. The above process was automated using Perl.

High resolution versions of all figures, associated details and Supplementary Information available at http://baderlab.org/Data/RoadsNotTaken